\documentclass[fleqn,twoside]{article}
\usepackage{espcrc2}

\usepackage{graphicx}
\usepackage[figuresright]{rotating}

\newcommand{\AmS}{{\protect\the\textfont2
  A\kern-.1667em\lower.5ex\hbox{M}\kern-.125emS}}

\title{Status of NEMO: results from the NEMO Phase-1 detector}

\author{C. Distefano\address[LNS]{Istituto Nazionale di Fisica Nucleare, Laboratori Nazionali del Sud, Via S. Sofia 62, 95123 Catania, Italy}, for the NEMO Collaboration}
       
\begin{document}

\begin{abstract}
The NEMO Collaboration installed an underwater detector including most of the critical elements of a possible km$^3$ neutrino telescope: a four-floor tower (called Mini-Tower) and a Junction Box, including the data transmission, the power distribution, the timing calibration and the acoustic positioning systems. These technical solutions will be evaluated, among others proposed for the construction of the km$^3$ detector, within the KM3NeT Consortium.
The main test of this test experiment was the validation of the proposed design solutions mentioned above. 
We present results of the analysis of data collected with the NEMO Mini-Tower. The position of PMTs is determined through the acoustic position system; signals detected with PMTs are used to reconstruct the tracks of atmospheric muons. The angular distribution of atmospheric muons was measured and results were compared with Monte Carlo simulations.
\end{abstract}

\maketitle

\section{INTRODUCTION}

Due to the expectations on neutrino fluxes from galactic
and extragalactic sources, mainly based on the measured
cosmic ray fluxes and the estimated fluxes from theoretical
models \cite{modelli}, the opening of the high-energy neutrino
astronomy era can only be made with detectors
of km$^3$ scale.

The activity of the NEMO Collaboration was
mainly focused on the search and characterization of an
optimal site for the detector installation and on the
development of key technologies for the km$^3$ underwater
telescope to be installed in the  Mediterranean Sea.

A deep sea site with optimal features in terms of depth
and water optical properties has been identified at a depth
of 3500 m about 80 km off-shore Capo Passero (Southern cape of Sicily).  A long
term monitoring of the site has been carried out \cite{sito2}. 

One of the efforts undertaken by the NEMO Collaboration
has also been the definition of a feasibility study of the
km$^3$ detector, which included the analysis of all the
construction and installation issues and the optimization
of the detector geometry by means of numerical simulations.

The technical solutions, proposed by the NEMO Collaboration, will be evaluated, among others proposed for the construction of the km$^3$ detector, within the KM3Net Consortium \cite{km3net}.

As an intermediate step towards the construction of the underwater km$^3$ detector and to ensure an adequate process of validation, the NEMO Collaboration built a technological demonstrator and installed off-shore the port of  Catania (Sicily). 
The project, called NEMO Phase-1, has allowed test and qualification of the key technological elements (mechanics, electronics, data transmission, power distribution, acoustic positioning and time calibration system) proposed for the km$^3$ detector
 \cite{Migneco08}. 
  After a brief description of the detector lay-out, we describe the atmospheric muon data analysis procedure and present the results. In particular, the atmospheric muon angular distribution was measured 
 and compared with Monte Carlo simulations.
 
 The NEMO Collaboration is also constructing an underwater infrastructure at the Capo Passero site (NEMO Phase-2).
 The main goal of this project is to finally validate the technologies proposed for the realization and
installation at the depths needed for the km$^3$ detector. 
The status of NEMO Phase-2 is also presented.

\section{THE NEMO PHASE-1 DETECTOR} 

\begin{figure}[htb]
\begin{center}
\includegraphics[height =4cm]{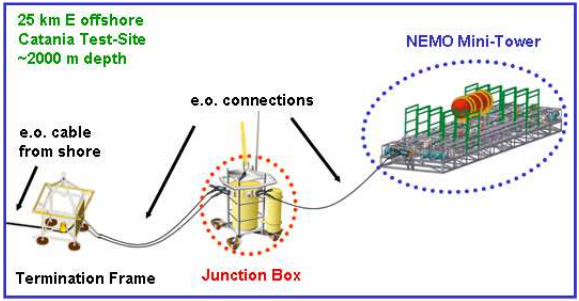}
\end{center}
\caption{Lay-out of the NEMO Phase-1 installation at the Catania TestSite.}
\label{fig:ts}
\end{figure}

The apparatus includes prototypes of the critical elements of the proposed km$^3$ detector \cite{Migneco08}: the Junction Box (JB) and four floor NEMO Tower (the Mini-Tower), as sketched in Figure \ref{fig:ts}.

\subsection{The Junction Box}
The JB is a key element of the detector. It
must provide connection between the main electro-optical
cable and the detector structures and has been
designed to host and protect from the effects of corrosion
and pressure the opto-electronic boards dedicated to the
distribution and the control of the power supply and
digitized signals.

The NEMO Phase-1 JB has been built following the
concept of double containment. Pressure resistant steel
vessels are hosted inside a large fiberglass container. This
last one is filled with silicon oil and pressure compensated.
This solution has the advantage to decouple the two
problems of pressure and corrosion resistance.

Moreover, all the electronics components that were
proven able to withstand high pressure were installed
directly in the oil bath.

\subsection{The Mini-Tower}
\label{sec:minitower}

The  Mini-Tower is a prototype of the NEMO Tower \cite{Migneco06}.
It is a three dimensional flexible structure
composed by a sequence of four floors 
interlinked by cables and anchored on the
seabed. The structure is kept vertical by appropriate
buoyancy on the top.

Each floor is made with
a 15 m long structure hosting two photomultipliers (PMTs) 
(one down-looking and one horizontally looking) at each
end (4 PMTs per storey). 
Besides, each floor is connected to the following one by
means of four ropes that are fastened in a way that forces
each floor to take an orientation perpendicular with respect
to the adjacent (top and bottom) ones. 
The floors are vertically spaced by 40 m.  An additional
spacing of 150 m is added at the base of the tower, between
the tower base and the lowermost floor to allow for a
sufficient water volume below the detector.

\begin{figure}[htb]
\begin{center}
\includegraphics[height =8cm]{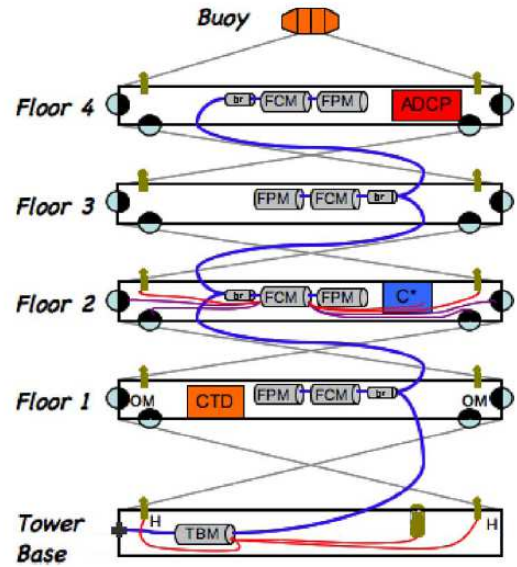}
\end{center}
\caption{Fully equipped NEMO Mini-Tower pictorial view (see text in sec. \ref{sec:minitower} and \ref{sec:daq}) for details.}
\label{fig:schema}
\end{figure}

In addition to the 16 PMTs the
instrumentation installed on the Mini-Tower includes several sensors for
calibration and environmental monitoring, such as 
oceanographic instrumentation to measure water current (ADCP), water
transparency (C*), sea water properties (CTD), and a pair of hydrophones (H)
for acoustic positioning. 
A scheme of the fully equipped Mini-Tower is shown in Fig. \ref{fig:schema}.

\section{DETECTOR OPERATION}
\label{sec:operations}

\begin{figure}[h]
\begin{center}
\includegraphics[height =5.5cm]{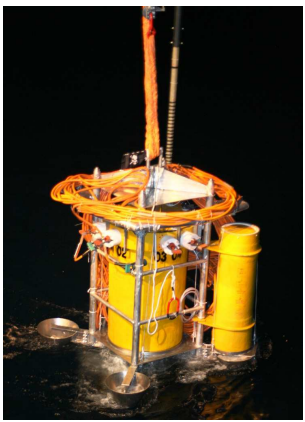}\vspace{0.5 cm}
\includegraphics[height =5cm]{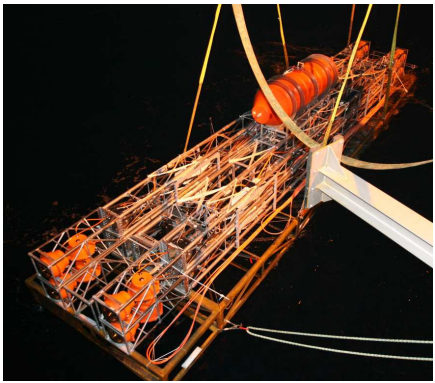}
\end{center}
\caption{Pictures of the JB (top) and of the Mini-Tower (bottom) deployment.}
\label{fig:deployment}
\end{figure}

The  NEMO Phase 1 detector was deployed in December 2006 (see Fig. \ref{fig:deployment}).
The apparatus was connected to the Underwater TestSite infrastructure of the Laboratori Nazionali del Sud, installed off shore Catania at a depth of about 2000 m.

The Mini-Tower deployment, connection and unfolding operations went smoothly. All active elements, such as PMTs, electronics, acoustic positioning, data transmission and acquisition, worked correctly.
 
 After 4 months, an attenuation in the optical fibers transmission was observed inside the JB. In May 2007, the JB was shut down because of a short-circuit. The JB was recovered in June 2007; it was repaired and re-installed
 in April 2008. It's working since then.
A poor manufacturing process caused a loss of buoyancy in the main buoy of the Mini-Tower and therefore a slow sinking of the whole tower.

Although these problems occurred during operations, the NEMO Phase-1 project successfully validated the key technologies proposed for the km$^3$ telescope, demonstrating also the NEMO Tower capability to detect and trace muons, as discussed in the following sections.

\section{ATMOSPHERIC MUON DATA ACQUISITION}

\subsection{The data acquisition system}
\label{sec:daq}

The PMT Front-End Module (FEM) was placed close to the PMT inside 
the glass sphere containing the optical module (OM).
The FEM main goal is to acquire the analog signals produced by the PMT,
encode and transmit these data to the Floor Control Module (FCM) (see Fig. \ref{fig:schema}).
The hit pulse is sampled by two 8-bits Fast Analog to Digital
Converters (Fast-ADC) running at 100 MHz but staggered
by 5 ns.

The PMT signals, produced by the FEM
boards, are collected by the FCM, packed together and transmitted
through the optical link; control data are received from
on-shore following the opposite direction. The on-shore host
machine, called Floor Control Module Interface (FCMI), can
be accessed through a Gigabit Ethernet (GbE) connection.
The floor control module (FCM), roughly placed at the floor center, is powered by the floor
power module (FPM) and connected to the Tower Base Module (TBM) through a fiber optic backbone.
The TBM is connected through an inter-link cable to the JB and therefore to the on-shore lab. 
A detailed description of the Mini-Tower data acquisition and transport systems is given in \cite{Ameli08}.

\subsection{The On-Line Trigger}

The PMT raw data were sent by the FCM boards to the On-Line Trigger.
The aim of this data processing was to select single time windows with a high probability to contain muon events.
In this way most of the optical background was rejected, strongly reducing the data to be recorded.

The On-Line Trigger algorithm was based on searching the so called Simple Coincidences (SCs) among the hits.
A SC is defined as a coincidence between 2 hits in 2 adjacent PMTs, placed at the same tower floor. The  coincidence 
time delay was set to $\Delta T_{SC} \le 20$ ns.

The trigger occurred when a SC was found. In this case,  the On-Line Trigger stored all the hits recorded in a time window centered around the SC
and long enough to contain the possible muon event. 
Two different values of the trigger time window were tested: 4 and 10 $\mu$s.
 
The event  detection rate at the On-Line Trigger level ranges between 1.5 and 2 kHz. This value is consistent with hit coincidences induced by the measured optical background rate of 75-80 kHz \cite{Amore08}.
The expected atmospheric muon trigger rate, evaluated from Monte Carlo simulations, is $\sim$1 Hz. The signal is dominated by the noise and an Off-Line Trigger is therefore mandatory.

\section{ATMOSPHERIC MUON DATA ANALYSIS}

\subsection{PMT data calibration}

Before atmospheric muon data analysis could be started, the recorded PMT hits had to be decompressed and calibrated \cite{Simeone08}. 
The hit wave-form is firstly re-sampled at 2 GHz (the ADC sampling is 200 MHz); the ADC channels are then decompressed and converted into amplitudes (in mV unit), using the decompression table generated during the FEM Boards characterization phase. 
The sample waveform rising edge is fitted with a sigmoid function and hit time is evaluated at the sigmoid inflection point.
Time offsets provided by the time calibration system are finally added.
At the end of the process the PMT hit waveform is reconstructed: the integral charge is determined with $\sigma\sim0.3$ pC and converted in units of p.e. taking into account that 1 p.e. = 8 pC.
The time is evaluated with an accuracy of $\sigma\sim1$ ns.

\subsection{The Off-Line Trigger}

After the calibration procedure, the hit time estimate is 5 times better than the raw data level.
For each event, the simple coincidences (SCs) were then re-calculated in order to reject the false SCs found by the On-Line Trigger.   
Besides, the new trigger seeds were calculated:

\begin{itemize}
\item Floor Coincidence (FC): a coincidence between 2 hits recorded at the opposite ends of a same storey ($\Delta T_{FC} \le 200$ ns);
\item Charge Shooting (CS): a hit exceeding a charge threshold of 2.5 p.e. 
\end{itemize}

The ensemble of all hits participating to the Off-Line Trigger seeds was then analyzed. In particular, for each one we calculated the number of the other hits
causality correlated  according to the condition: 
\begin{equation}
|dt|<dr/v + 20 \hbox{ ns},
\label{eq:causal}
\end{equation}
where $|dt|$ is the absolute value of the time delay between the hits, $dr$ is the distance between the PMTs where the hits are detected, $v$ is the group velocity of light in seawater.

The maximum number of causality relations $N_{Caus}$ found in each event is then used to reject the background. In particular only the events having  $N_{Caus}\ge4$ were considered in the following steps.

\subsection{The Causality Filter}

Before trying any track reconstruction, it is mandatory to reject the background hits present inside the muon event.

In order to reduce the number of hits due to background the first step consists in the rejection of hits with amplitude
smaller than 0.5 p.e. then a causality filter with respect to a reference hit is applied. 
The causality filter application proceeded in the following way.

The $N$ hits forming the event were sorted by time and the local frequency was calculated as:
\begin{equation}
f=\frac{N}{T_N-T_1},
\end{equation}
where $T_1$ and $T_N$ are respectively the occurrence time of the oldest and the youngest hits.

For each group of $n$ ($n=5$) consecutive hits, we calculated the Poisson probability to detect $n$ background hits for an expected value of
\begin{equation}
n_{exp}=f\cdot\Delta T_g,
\end{equation}
 where $\Delta T_g$ is the time interval in which the $n$ hits were detected.
The hit group with the minimum probability is likely to contain muon hits, instead of uncorrelated background hits.   

The causality filter is then applied with respect to all hits in the group with the minimum probability.
In particular for each reference hit, the number of hits among the $N$ forming the event and selected by the 
same condition in Eq. \ref{eq:causal} are counted. Among the $n$ cases, the one that preserves the largest number of hits
is chosen. 

\subsection{Muon track reconstruction}

The hits surviving the causality filter were used to reconstruct the atmospheric muon tracks.
The track reconstruction strategy used in this analysis is a robust track fitting procedure based on a maximum likelihood method.
The algorithm takes into account the \v{C}erenkov light features and the possible presence of unrejected background hits \cite{Heijboer03,Distefano07}.
During the reconstruction procedure, the PMTs positions, reconstructed using acoustic positioning system data, were considered.

\subsection{Results}

A sample of data, recorded on 23$^{\hbox{rd}}$-24$^{\hbox{th}}$ January 2007 when the tower was completely unfolded and corresponding to a livetime of 11.3 hours,
was analyzed. A total of 3049 atmospheric muon events was reconstructed, corresponding to a mean reconstruction rate of 0.075 Hz, and
their angular distribution was measured.

For comparison, a Monte Carlo simulation of the detector response to atmospheric muons
was carried out. A total of $4\cdot 10^7$ atmospheric muon events were simulated with MuPage \cite{Carminati08}, 
corresponding to a livetime of 11.3 hours. The detector response was simulated taking into account the light absorption length spectrum measured at the TestSite and 
the optical background evaluated from the measured PMT data.
The Mini-Tower DAQ electronics and the On-Line Trigger were simulated.
The detector geometry was simulated using the PMTs positions reconstructed using the acoustic positioning system data.

\begin{figure}[htb]
\begin{center}
\includegraphics[width =7cm]{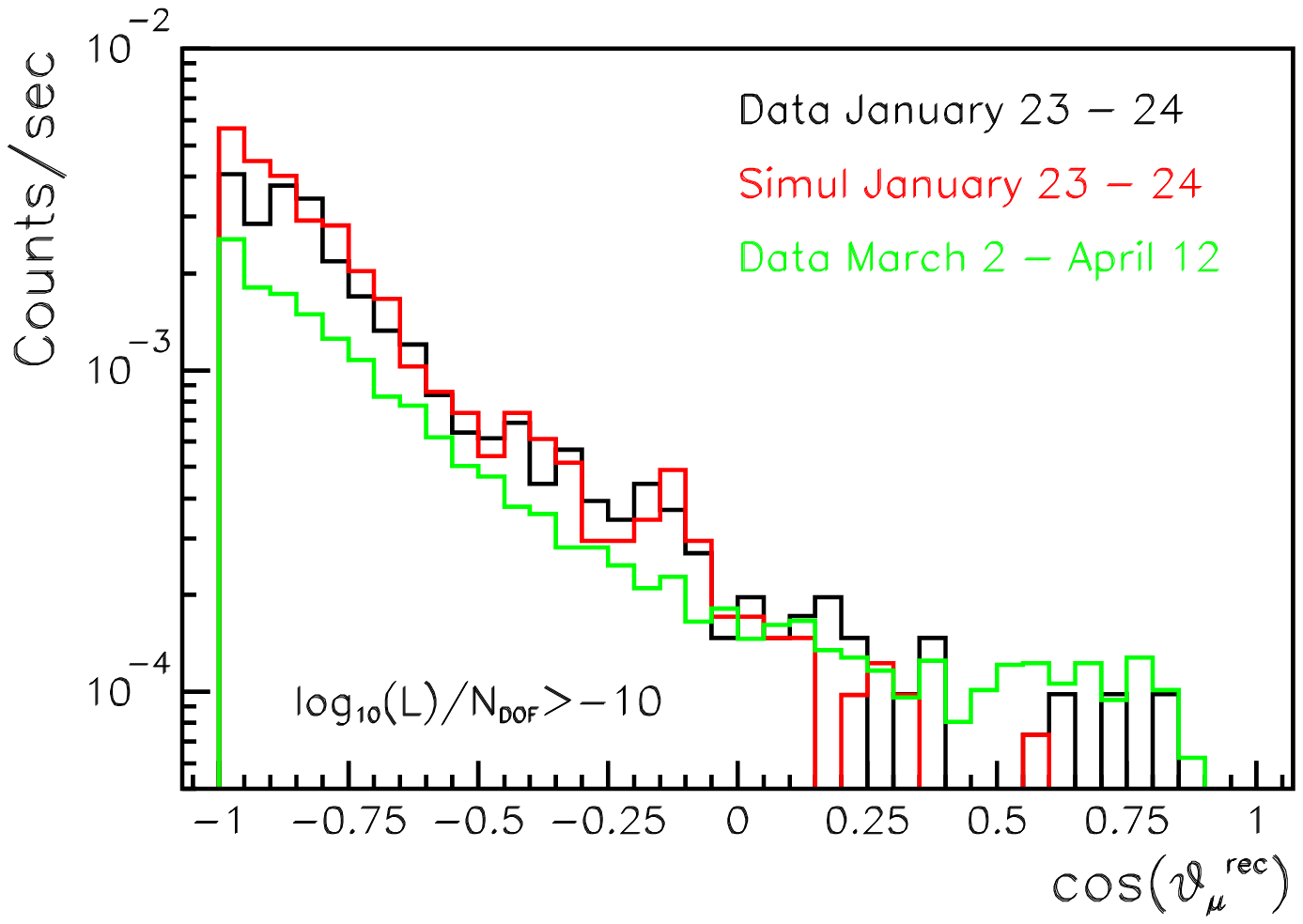}
\includegraphics[width =7cm]{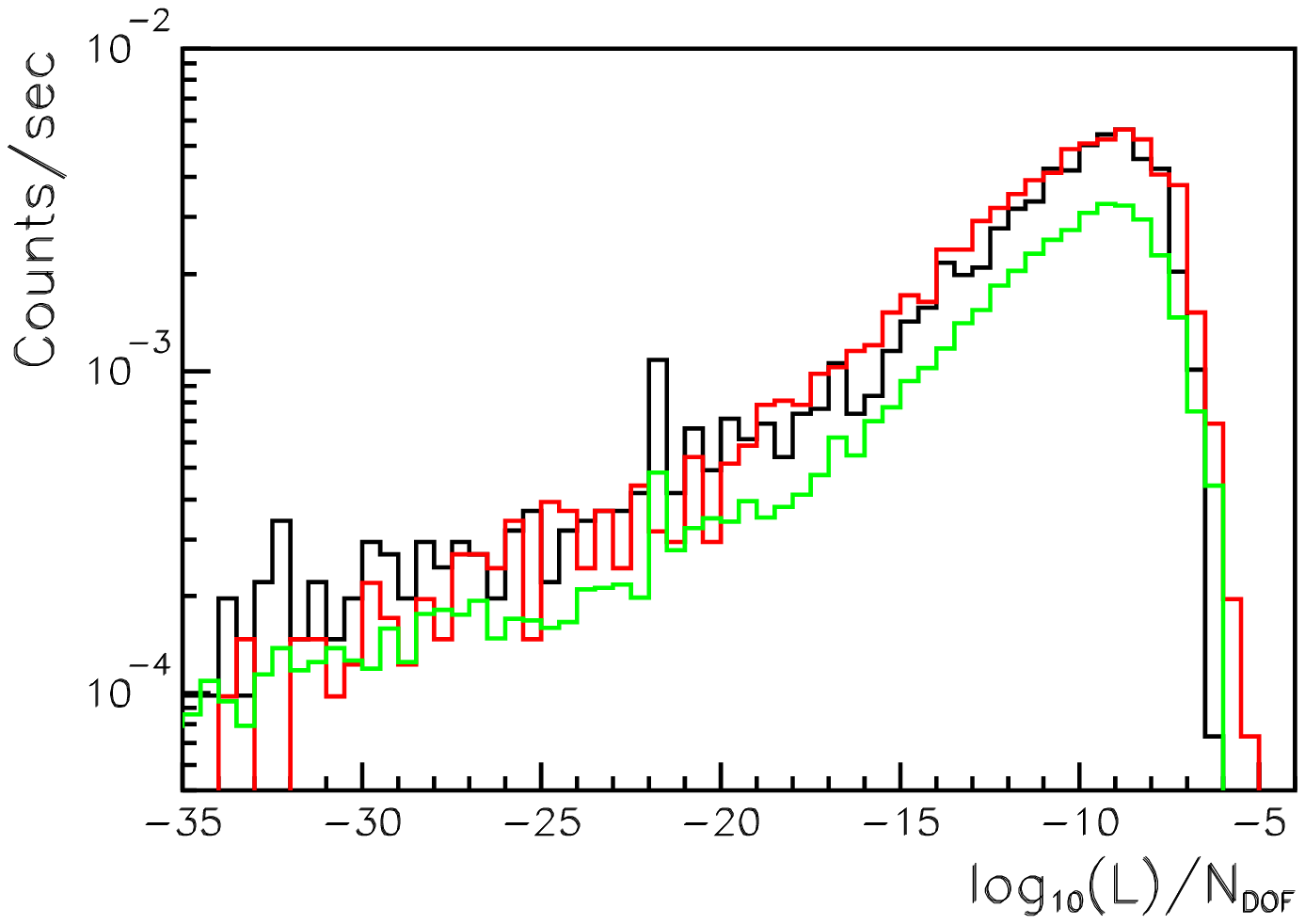}
\end{center}
\caption{Angular distribution of reconstructed muon tracks applying a likelihood quality cut (top panel) and the reconstruction likelihood spectrum (bottom panel).  }
\label{fig:muatm}
\end{figure}

The angular distribution of reconstructed muon tracks, detected during the  23$^{\hbox{rd}}$-24$^{\hbox{th}}$ January 2007 period,
is shown in Fig. \ref{fig:muatm} together with the spectrum of the reconstruction likelihood. 
The figure reports also results from Monte Carlo simulations carried out for the same period, showing an excellent agreement. 

Data recorded during the period between 2$^{\hbox{nd}}$  March  and 12$^{\hbox{th}}$ April 2007 were also analyzed. 
At that time, the two lowest floors of the tower were already laying on the seabed.  
The corresponding livetime is
174.1 hours and the total number of reconstructed atmospheric muons is 27699 (reconstruction rate 0.044 Hz).
The lower rate of reconstructed tracks is due to the smaller number of PMTs participating to the muon detection caused by the slow Mini-Tower sinking (see sec. \ref{sec:operations}).

\section{NEMO PHASE-2}

Although the Phase-1 project provided a fundamental
test of the technologies proposed for the realization and
installation of the detector, these must be finally validated
at the depths needed for the km$^3$ detector. For these
motivations the realization of an infrastructure on the site
of Capo Passero has been undertaken. It consists of a
100 km cable, linking the 3500 m deep sea site to the shore,
a shore station, located inside the harbor area of Portopalo
of Capo Passero, and the underwater infrastructures
needed to connect prototypes of the km$^3$ detector.
At the same time a fully equipped 16 storey detection
tower is under construction and will be installed on the
Capo Passero site. With the completion of this project,
foreseen by the spring of 2009, it will be possible to perform a
full test at 3500 m of the deployment and connection
procedures and at the same time set up a continuous long
term on-line monitoring of the site properties (light
transparency, optical background, water currents, ...)
whose knowledge is essential for the installation of the full
km$^3$ detector.

Due to the longer cable needed, with respect to the
Phase-1 project, the DC solution was chosen for the
electro-optical cable power feeding. The main cable,
manufactured by Alcatel, carries a single electrical conductor,
that can be operated at 10 kV DC allowing a power
transport of more than 50 kW, and 20 single mode optical
fibres for data transmission. The DC/DC converter
will be realized by Alcatel and will convert the high voltage
coming from the shore into 400 V.
The cable has been laid in July 2007. The cable deep sea
termination, that includes the 10 kW DC/DC converter
system, is presently under realization and will be deployed
in the beginning of 2009.

\section{CONCLUSIONS}

The activities of the NEMO Collaboration progressed in the past two years with the realization and installation of the Phase-1 apparatus. With this apparatus it has been possible to test in deep sea the main technological solutions developed by the collaboration for the construction of a km$^3$ scale underwater neutrino telescope \cite{sito2}. In particular the angular distribution of atmospheric muons was measured and results were compared with Monte Carlo simulations, finding an reasonable agreement. 
A Phase-2 project, which aims at the realization of a new infrastructure on the deep-sea site of Capo Passero at 3500 m depth, is presently progressing. 
After a careful revision of its design, following the experience gained with the Phase-1 project, the construction of a fully equipped 16 storey tower is under way. The tower will be installed and connected in spring of 2009.


\begin{thebibliography}{9}
\bibitem{modelli} J.G. Learned, K. Mannheim, Ann. Rev. Nucl. Part. Sci. 50 (2000) 679.
\bibitem{sito2} G. Riccobene et al., Astrop. Phys. 27 (2007) 1.
\bibitem{km3net}  KM3NeT web page, http://www.km3net.org.
\bibitem{Migneco08} E. Migneco et al., NIM A 588 (2008) 111. 
\bibitem{Migneco06} E. Migneco et al., NIM A 567 (2006) 444.
\bibitem{Ameli08} F. Ameli et al., IEEE Trans. Nucl. Sci. 55 (2008) 233.
\bibitem{Amore08} I. Amore for the NEMO Coll., NIM A (2008), in press (arXiv:0810.3119 [astro-ph]).
\bibitem{Simeone08} F. Simeone for the NEMO Coll., NIM A 588 (2008) 119.
\bibitem{Heijboer03} A. Heijboer, 2004, {\it Track reconstruction and point source searches with Antares}, PhD dissertation, Universiteit van Amsterdam, Amsterdam,  The Netherlands (http://antares.in2p3.fr/).
\bibitem{Distefano07} S. Aiello et al.,  Astrop. Phys. 28 (2007) 1.
\bibitem{Carminati08} G. Carminati et al., Computer Physics Communications 179 (2008) 915.
\end{thebibliography}
\end{document}